\documentstyle[aps,twocolumn,prb,epsfig,amstex]{revtex}
\title{Stark effect in colloidal indium arsenide nanocrystal quantum 
       dots: consequences for wave function mapping experiments}
\author{Michael Tews and Daniela Pfannkuche}
\address{I. Institute of Theoretical Physics, University of Hamburg,
         Jungiusstr. 9, 20355 Hamburg, Germany}
\begin{document}
\maketitle

%
%----------------------------------------------------------------

\begin{abstract}
  The influence of the tip-substrate bias induced electric field in a 
  scanning tunneling spectroscopy experiment on colloidal InAs 
  nanocrystals has been studied. Calculating the Stark induced 
  splitting of the degenerate $1P_e$ state perturbatively within a 
  particle-in-a-sphere model, revealed a possible explanation of 
  recently published experimental wave function mapping data
  by Millo et al. \cite{millo:06:2001}. 
\end{abstract}


%----------------------------------------------------------------
\narrowtext
%%%%%%%%%%%%%%%%%%%%%%%%%%%%%%%%%%%%%%%%%%%%%%%%%%%%%%%%%%%%%%%%%%%%%%%%
\section{Introduction}
%%%%%%%%%%%%%%%%%%%%%%%%%%%%%%%%%%%%%%%%%%%%%%%%%%%%%%%%%%%%%%%%%%%%%%%%
  Mapping of the electronic wave functions in semiconductor quantum dots
  (QD) has recently become possible using various experimental 
  techniques \cite{millo:06:2001,grandidier:07:2000,vdovin:10:2000}.
  Knowing the actual shape of the electronic densities contributes to a
  better understanding of the QD electronic structure. This knowledge is 
  crucial with respect to the possible importance of semiconductor QDs 
  as the ultimate building blocks of optoelectronic and nanoelectronic
  devices. 

  Most recent scanning tunneling microscopy (STM) measurements 
  \cite{millo:06:2001} also allow a wave function mapping of colloidal
  nanocrystals. The used InAs/ZnSe core/shell structures are spherical
  in shape resulting in atomic-like symmetries and degeneracies of the 
  conduction band (CB) states \cite{banin:08:1999,alperson:09:1999}. 
  This leads, for example, to a six fold degenerate first excited state
  (here after referred to as the $1P_e$ state, where the subscription 
  $e$ denotes an electron rather than a hole state) such that
  a mapping of the $1P_e$ wave function should reveal the spherical
  superposition of all degenerate states. In contrast to this simple 
  prediction the experimental data \cite{millo:06:2001} show a torus
  like $p_{x^2+y^2}$ symmetry.

  We present a calculation of the CB states within a 
  particle-in-a-sphere model taking into account the electric field 
  due to the applied STM voltage. The resulting quantum-confined Stark
  effect was studied earlier in three dimensional QDs 
  \cite{wen:08:1995,li:12:2000}. In this work we concentrate on the
  Stark effect induced degeneracy lifting of the first excited state
  which provides an explanation for the observed $p_{x^2+y^2}$ wave
  function symmetry \cite{millo:06:2001} at appropriate STM voltages.

%%%%%%%%%%%%%%%%%%%%%%%%%%%%%%%%%%%%%%%%%%%%%%%%%%%%%%%%%%%%%%%%%%%%%%%%
\section{Model}
\begin{figure}[htb]
    \epsfig{file=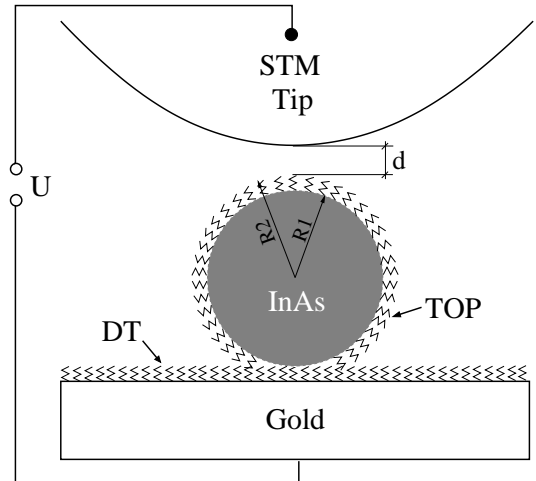,width=70mm}
    \caption{Scanning tunneling spectroscopy of a single InAs nanocrystal.
             The InAs nanocrystal with a typical radius of a few
             nanometers are linked to a gold substrate by hexane dithiol
             molecules (DT). Trioctylphosphin (TOP) molecules form
             a ligand shell around the nanocrystal. At $4.2K$ the tunnel
             current is measured as a function of the applied voltage U
             between tip and substrate.}
    \label{fig_experiment}
  \end{figure}

%%%%%%%%%%%%%%%%%%%%%%%%%%%%%%%%%%%%%%%%%%%%%%%%%%%%%%%%%%%%%%%%%%%%%%%%
  A sketch of the experimental setup in a scanning tunneling 
  spectroscopy (STS) experiment on colloidal nanocrystal quantum dots is 
  shown in Fig. \ref{fig_experiment}. To obtain a tunnel spectrum the 
  tip is positioned above a single nanocrystal attached to the substrate
  via hexane dithiol molecules. Keeping the tip-crystal distance constant
  the differential conductance as a function of applied voltage shows 
  sharp peaks \cite{banin:08:1999,millo:06:2000,bakkers:09:2000}. 
  For a detailed understanding of the experimental data it is thus 
  crucial to know the discrete energetic spectrum of the QD since the 
  obtained peak positions $U_{peak}$ are directly related to the 
  electronic dot structure \cite{bakkers:09:2000}:
  \begin{eqnarray}
    eU_{peak}(N,N+1) = \gamma [E(N+1,\mu)-E(N,\nu)] \\
    E(N,\mu=\{n_i\}) = \sum_{i} n_i \epsilon_i + V_Q^{tot}(\{n_i\}) 
    \label{Eq_TotalEnergy}
  \end{eqnarray}
  where the pre-factor $\gamma$ depends on the capacitive electrostatic
  geometry. For a very asymmetric tip-dot dot-substrate capacity
  distribution $\gamma$ is close to one \cite{bakkers:09:2000}.
  The total energy $E(N,\{n_i\})$ of a dot with $N=\sum_i n_i$ electrons,
  where $n_i$ denotes the occupation number of state $i$, is written as a
  sum of
  occupied single particle levels with energies $\epsilon_i$ and the total
  charging energy $V_Q^{tot}$. Hereby $V_Q^{tot}$ includes both, the direct
  Coulomb interaction and all correlations.

  In the STS setup of Fig. \ref{fig_experiment} with voltages up to
  $U \approx 2V$ \cite{millo:06:2001,banin:08:1999} applied on a
  tip-substrate distance of a few nanometers, the QD is exposed
  to a considerable electric field. While a STM tip has to terminate
  in a single atom in order to achieve atomic resolution, the macroscopic
  tip size is usually about one order in magnitude bigger than the here
  studied nanocrystals \cite{wiesendanger:1994}. Other than a macroscopic
  metallic tip, a single terminating atom is not able to substantially
  focus the electric field. Over the nanocrystal size we can therefore
  assume the field between tip and substrate to be homogeneous in absence
  of the QD.
  Since the considered InAs nanocrystals are surrounded by ligands with a quite
  different relative dielectric constant compared to InAs, we modeled
  the QD as a jacketed dielectric sphere. Extending the text book
  calculation of the electric potential $\phi_{InAs}$ inside a
  dielectric sphere placed in a homogeneous field ${\cal E}_{hom}$
  \cite{griffiths_edyn:1998} to such a structure leads in spherical
  coordinates to:
  \begin{eqnarray}
    \lefteqn{\phi_{InAs}(r,\theta) =} \notag \\
    & & \frac{9\epsilon_{2} {\cal E}_{hom} r cos\theta}
    {2\epsilon_{2}^2+\epsilon_{1} \epsilon_{2}+4\epsilon_{2}+
    2\epsilon_{1}+2(\frac{R_{1}}{R_{2}})^3[\epsilon_{1}\epsilon_{2}
     +\epsilon_{2}-\epsilon_{1} - \epsilon_{2}^2]}
    \label{eq_corePotential}
  \end{eqnarray}
  with $\epsilon_{1}=15.15$ \cite{madelung:1996} and $\epsilon_{2}=2.1$
  \cite{kim:06:2001} being the relative dielectric constants of InAs  
  and the ligands respectively. $R_1$ is the InAs core radius and $R_2$
  the total radius including the ligand shell 
  (see Fig. \ref{fig_experiment}). As in the case of a dielectric
  sphere without a shell \cite{griffiths_edyn:1998} the core potential 
  $\phi_{InAs}$ is still the potential of a homogeneous field. The
  field ${\cal E}_{hom}$  occurring in (\ref{eq_corePotential}) is
  not directly accessible. It is obtained by equating the voltage $U$
  with the potential drop between tip and substrate of the 
  inhomogeneous field outside the QD. 

  The single particle spectrum of conduction band (CB) states was
  calculated using a particle-in-a-sphere model with a finite potential
  well \cite{brus:05:1984}. Such a spherical confinement leads to size 
  quantization comparable to the bulk energy gap of InAs. Hence 
  non-parabolicity effects of the CB have to be taken into account.
  We use an energy dependent effective mass approach \cite{burt:03:1992}
  with
  \begin{equation}
    m^*(E) = m^*(0) [1+{E}/{E_g}] 
  \end{equation}
  where $m^*(0)$ is the bottom CB effective mass and $E_g$ the bulk
  energy gap. The validity of this approach has been checked by Bryant
  \cite{note:2} for a dot radius of 3.2 nm by comparison to a 
  8 band calculation including CB-valence band coupling.

  Using the obtained electric potential inside the QD our model
  Hamiltonian is:
  \begin{equation}
    H = -\frac{\hbar^2}{2m^*(E)}\nabla^2+V(r)-e\phi_{InAs}(r,\theta)
    \label{eq_hamiltonian}
  \end{equation}
  with $V(r)$ being the potential step with step height corresponding
  to the work function of InAs towards air. The ligand shell is not 
  included in the potential $V(r)$.

  For calculating the QD states in the field we treated the
  electric potential perturbatively. The energies are
  obtained using second order perturbation theory \cite{note:1}.

%%%%%%%%%%%%%%%%%%%%%%%%%%%%%%%%%%%%%%%%%%%%%%%%%%%%%%%%%%%%%%%%%%%%%%%
\section{Results}
\begin{figure}[htb]
    \epsfig{file=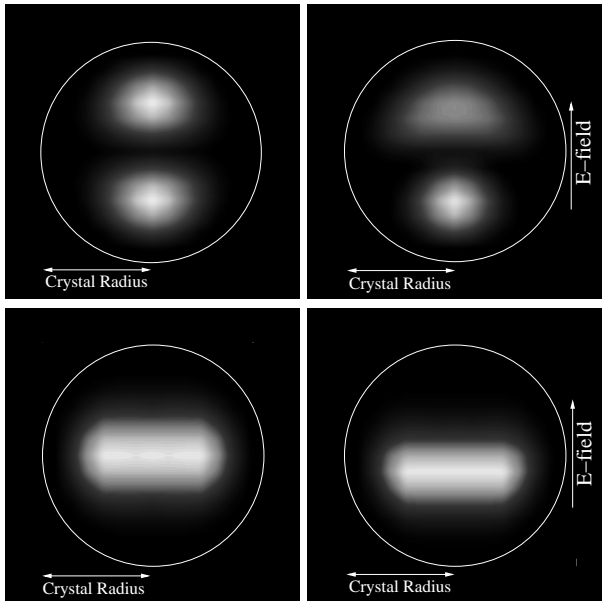,width=80mm}
    \caption{Electronic density of the $1P_e$ envelope wave functions
             calculated within a particle-in-a-sphere model. 
             The densities illustrate qualitatively the different Stark 
             effect on the $1P_e$ state, although they were calculated
             in an infinite spherical well neglecting the 
             non-parabolicity of the conduction band. Top and
             bottom left show the $1P_e (m\pm1)$ and $1P_e (m=0)$
             densities without the presence of an electric field 
             respectively. Whereas top right and bottom right show the
             corresponding densities in a homogeneous field of 
             $0.18$ V/nm} 
    \label{fig_densitysideview}
  \end{figure}

\begin{figure}[htb]
     \epsfig{file=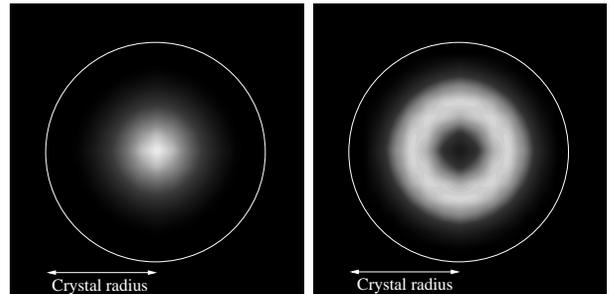,width=80mm}
     \caption{The plots above show the electronic density of Fig.
             \ref{fig_densitysideview} viewed along the applied 
             electric field. The $1P_e (m=0)$ density is shown
             left whereas on the right side the superposition of
             $1P_e (m=1)$ and $1P_e (m=-1)$ is plotted.}
     \label{fig_densitytopview}
  \end{figure}

%%%%%%%%%%%%%%%%%%%%%%%%%%%%%%%%%%%%%%%%%%%%%%%%%%%%%%%%%%%%%%%%%%%%%%%
  Without an electric field the Hamiltonian (\ref{eq_hamiltonian}) 
  separates in an angular and a radial part. The angular Schr\"odinger
  equation is solved by spherical harmonics, and the radial
  Schr\"odinger equation by spherical Bessel functions $j_l$ inside 
  the well and spherical Hankel functions $h_l$ outside 
  \cite{schiff:1993} the well. 
  The continuity conditions at the potential step lead to a set of
  transcendental equation determining the energy levels:
  \begin{eqnarray}
    \lefteqn{ \alpha h_l(i \beta R_1) \left[ l j_{l-1}(\alpha R_1)
    -(l+1)j_{l+1}(\alpha R_1) \right] =} \notag \\
    & &  i \beta j_{l}(\alpha R_1) \left[ l h_{l-1}(i\beta R_1)
    -(l+1)h_{l+1}(i \beta R_1) \right]
  \end{eqnarray}
  with $\alpha = \sqrt{2m^* E}/\hbar$ and 
  $\beta=\sqrt{2m^*(V-E)}/\hbar$. The numerical solution of the first
  two energy levels for a dot radius
  of 3.2 nm is shown in Tab. \ref{tab_energies}. Other than in
  hydrogen the allowed orbital quantum numbers are not restricted by
  the principal quantum number. Hence the first exited state has
  the quantum numbers $n=1$ and 
  $l=1$. Owing to the spherical harmonics this state is 3-fold
  degenerate in the three magnetic quantum numbers $m=-1,0$ and $1$ 
  (see Fig. \ref{fig_densitysideview} and \ref{fig_densitytopview}). 
  These unperturbed wave functions of (\ref{eq_hamiltonian}) are now 
  used to calculate the effect of the electric potential,
  $\phi_{InAs}$ using second order perturbation theory.

  \begin{figure}[htb]
    \center
    \epsfig{file=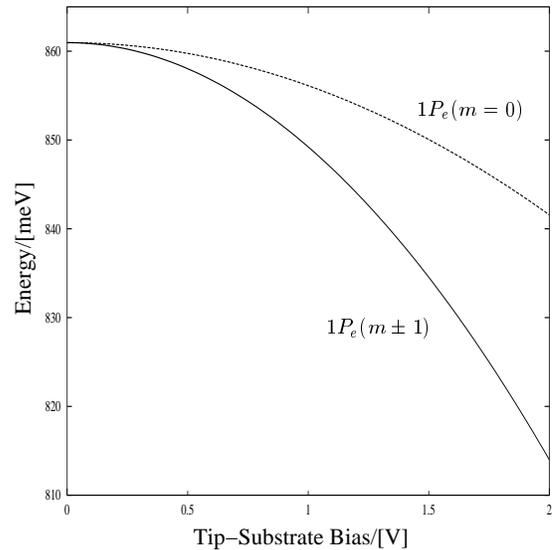,width=75mm}
    \caption{Stark splitting of the $1P_e$ level in a 3.2 nm radius 
             InAs nanocrystal as a function of the applied 
             tip-substrate voltage. For calculating the QD 
             potential the ligand shell was taken to be 
             $R_2-R_1=0.5$ nm thick whereas the tip-shell distance 
             was taken as another $d=0.5$ nm (see Fig.
              \ref{fig_experiment}). $3eV$ was used for the depth of
             the confining
             potential as obtained by \cite{williamson:06:1999} using a
             pseudopotential approach. To get the energy dependent
             mass, the InAs bulk energy gap $E_g=0.42eV$ 
             \cite{madelung:1996} and the effective mass 
             $m^*(0)=0.0239$ \cite{madelung:1996} in units of the
             free electron mass, were used.}
    \label{fig_starkeffect}
  \end{figure}   

  Although the $1S_e$ state also shows a Stark effect we concentrate 
  here on the first exited $1P_e$ state, with regard to the 
  experimental results obtained by Millo et. al. \cite{millo:06:2001}. 
  In contrast to hydrogen, the first excited state does not show a 
  linear Stark effect, owing to the lack of s-p degeneracy in such a
  spherical well. The $1P_e$ degeneracy is lifted by the quadratic 
  Stark effect, such that the energy of the  $1P_e(m\pm 1)$ wave 
  functions oriented perpendicular to the field (bottom row in Fig.
  \ref{fig_densitysideview}) are lowered compared to
  the $1P_e(m = 0)$ wave function oriented along the field (top row in 
  Fig. \ref{fig_densitysideview}). This behavior can be qualitatively
  understood by looking at the electronic densities.
  In an electric field energy is gained by moving the electronic 
  densities in field direction. The $1P_e(m\pm1)$ densities
  shown in the bottom row of Fig. \ref{fig_densitysideview} can move
  quite freely in that direction with only a small increase in confining 
  energy. Hence a strong dipole moment is induced leading to a pronounced 
  Stark effect. On the other hand the cost in potential energy for the
  $1P_e(m=0)$ density by moving into the same direction is much higher
  due to the close confinement potential step. This leads
  to a smaller induced dipole moment as can be seen in the top row 
  of Fig. \ref{fig_densitysideview}, and therefore to a smaller 
  quadratic Stark effect.

  The quantitative tip-substrate voltage dependence of the $1P_e$
  energy level is shown in Fig. \ref{fig_starkeffect} for a QD radius
  of $R_1=3.2$ nm.  At an applied voltage of 1.4 V as used in 
  experiment \cite{millo:06:2001} the energetic difference between
  $1P_e(m=0)$ and $1P_e(m\pm1)$ is about 15 meV. This Stark induced 
  energy splitting is experimentally resolvable and
  allows at an appropriate voltage tunneling into the energetically lower
  $1P_e(m\pm1)$ states without tunneling into the $1P_e(m=0)$ state. 
  The qualitatively different density distributions of those split
  states shown in Fig. \ref{fig_densitytopview} are observed in
  a STM experiment by Millo et. al. \cite{millo:06:2001}.  
  Therefore the obtained Stark induced degeneracy lifting of the 
  first excited $1P_e$ state serves as a possible explanation for
  experimentally observed mapping of the $1P_e(m\pm1)$ wave
  functions only (see discussion). 
  
%%%%%%%%%%%%%%%%%%%%%%%%%%%%%%%%%%%%%%%%%%%%%%%%%%%%%%%%%%%%%%%%%%%%%%%
\section{Discussion}
%%%%%%%%%%%%%%%%%%%%%%%%%%%%%%%%%%%%%%%%%%%%%%%%%%%%%%%%%%%%%%%%%%%%%%%
  As shown in this publication the first $1P_e$ electron will occupy
  one of the energetically lower $1P_e(m\pm1)$ states.
  The second electron is then going to occupy the other $1P_e(m\mp1)$ 
  state. This leads to a torus like electron density seen in a 
  wave function mapping experiment. Due to the exchange interaction 
  the spins of both electrons line up. 

\begin{figure}[htb]
    \epsfig{file=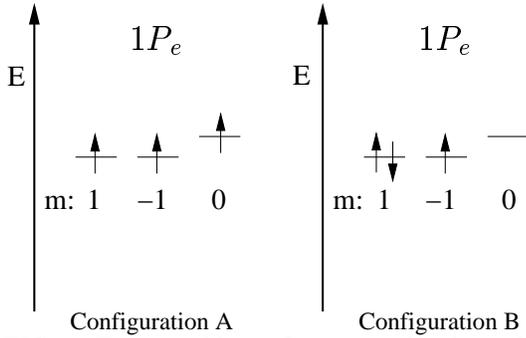,width=70mm}
    \caption{Two possible configurations for three electrons in
            the $1P_e$ states.} 
    \label{fig_configurations}
  \end{figure}

  If all $1P_e$ states had been degenerate the third electron would
  have followed Hund's rule by occupying the $1P_e(m=0)$ state with same
  spin orientation as the first two electrons. This configuration shown on
  the left of Fig. \ref{fig_configurations} would have led to a spherical
  electronic density. But due to the Stark effect the $1P_e$ 
  degeneracy is now lifted such that configuration A competes with 
  configuration B shown on the right of Fig. \ref{fig_configurations}.
  This configuration would lead to a torus like electronic density since
  no electron is occupying the $1P_e(m=0)$ state. 
  
  In experiment \cite{millo:06:2001} a torus like electron density was
  found with three electrons occupying the $1P_e$ state, suggesting
  that configuration B has a smaller total energy. 
  
  Configuration A has to pay extra energy for the third electron due to
  the energetically higher $1P_e(m=0)$ level. On the other hand all spins 
  are lined up such that this configuration takes advantage of the full
  exchange energy. In  configuration B the situation is just opposite.
  While the third electron occupies the energetically lower $1P_e(m=1)$ 
  level this configuration lacks the exchange energy of the third 
  electron due to its reversed spin.

  To find out which configuration has the lower energy, we calculate
  the contribution of the first order Coulomb interaction to the total
  energy. Using the  wave function of a infinite spherical well we
  estimate that the exchange energy of configuration B is
  approximately 13 meV higher than in configuration A. This is 2 meV
  smaller than the Stark shift of A and therefore configuration B is
  favored, in agreement with experiment.

  The first order Coulomb contribution for two electrons
  in the $1S_e$ state yields a charging energy of 56 meV which is about
  half of the experimental value \cite{banin:08:1999}. One
  reason for this small value is the neglected interaction 
  with the polarization charges at the QD's surface. These charges 
  induced by the electrons inside the QD are expected to be 
  large due to the high dielectric mismatch at the surface.
  While this interaction enhances the charging energy, it has only a 
  small effect on the exchange energy \cite{goldoni:06:2000}.

  Although our calculations explain the experimental data, the close
  competion between Stark effect and Coulomb interaction requires 
  a more thorough investigation. On the one hand the actual electronic
  structure of the InAs/ZnSe nanocrystal,
  and on the other hand the influence of correlations on the Coulomb
  energy in these few electron systems has to be taken into account.

%%%%%%%%%%%%%%%%%%%%%%%%%%%%%%%%%%%%%%%%%%%%%%%%%%%%%%%%%%%%%%%%%%%%%%%
\section{Conclusion}
%%%%%%%%%%%%%%%%%%%%%%%%%%%%%%%%%%%%%%%%%%%%%%%%%%%%%%%%%%%%%%%%%%%%%%%
  We calculated the Stark effect perturbatively on the first excited CB 
  state in an InAs nanocrystal within a particle-in-sphere model. An
  energy dependent effective mass was used to account for the 
  non-parabolicity in the CB. The obtained degeneracy lifting of the
  first excited $1P_e$
  state provides a possible explanation for experimentally observed 
  mapping of $1P_e(m\pm1)$ wave functions without $1P_e(m=0)$ mixing.

%%%%%%%%%%%%%%%%%%%%%%%%%%%%%%%%%%%%%%%%%%%%%%%%%%%%%%%%%%%%%%%%%%%%%%%
\section*{Acknowledgments}
%%%%%%%%%%%%%%%%%%%%%%%%%%%%%%%%%%%%%%%%%%%%%%%%%%%%%%%%%%%%%%%%%%%%%%%
  The authors gratefully acknowledge valuable discussions with
  Markus Morgenstern, Theophilos Maltezopoulos and Uri Banin. We
  would also like to thank Garnett Bryant for kindly providing
  extended band calculations on InAs nanocrystals. 
  This work was supported by the DFG through SFB 1641 and GrK 32048.
  
\bibliographystyle{prsty}
\bibliography{biblio,QDStarkBib}

\begin{thebibliography}{10}
\bibitem{note:1}
For the infinite confinement potential the 2nd order perturbation theory has
  been compared to the results of a full diagonalization. The relative error of
  the perturbative calculation versus the full diagonalization is less than
  $10^{-3}$ for voltages up to $U=2$V.
\end{thebibliography}

  \begin{table}
    \caption{This table shows the first two energy levels in an 
             $R_1=3.2$ nm InAs QD without an electric field.
             The energy difference between the ground and first excited
             states is in good agreement with experiment
             \cite{banin:08:1999}.}
    \label{tab_energies}
    \begin{tabular}{ccc}
      State &  Energy in meV & $E(1P_e)-E(1S_e)$\\ \tableline
      $1S_e$ &  542 &  \\ 
      $1P_e$ &  861 & 319
    \end{tabular}
  \end{table}
\end{document}